\documentclass[pre,aps,twocolumn,superscriptaddress,balancelastpage]{revtex4-1}

\usepackage{latexsym,amssymb,bm,bbold,amsfonts,amstext,graphicx,bbm,bm,relsize,dsfont,times}
\usepackage[dvipsnames]{xcolor}
\usepackage{enumerate}

\usepackage{amsmath}
\usepackage{amsthm}
\usepackage[unicode=true, breaklinks=false, pdfborder={0 0 1}, backref=false, colorlinks=true, linkcolor=blue, citecolor=blue]{hyperref}
\usepackage{subfiles}
\usepackage{tcolorbox}
\usepackage{tikz}
\usepackage{appendix}
\usepackage{comment}

\begin{document}

\def\be{\begin{equation}}
\def\ee{\end{equation}}
\def\bea{\begin{eqnarray}}
\def\eea{\end{eqnarray}}

\title{Irreversibility mitigation in unital non-Markovian quantum evolutions}

\author{Stefano Gherardini}
\email{gherardini@lens.unifi.it}
\affiliation{\mbox{Department of Physics and Astronomy \& LENS, University of Florence,} via G. Sansone 1, I-50019 Sesto Fiorentino, Italy.}
\affiliation{\mbox{SISSA}, via Bonomea 265, I-34136 Trieste, Italy}.

\author{Stefano Marcantoni}

\email{stefano.marcantoni@nottingham.ac.uk}
\affiliation{Dipartimento di Fisica, Universit\`a di Trieste \& \mbox{INFN}, Sezione di Trieste, I-34151 Trieste, Italy.}
\affiliation{School of Physics \& Astronomy, University of Nottingham, Nottingham NG7 2RD, UK}
\affiliation{Centre for the Mathematics and Theoretical Physics of Quantum Non-Equilibrium Systems, University of Nottingham, Nottingham, NG7 2RD, UK}

\author{Filippo Caruso}
\email{filippo.caruso@unifi.it}
\affiliation{\mbox{Department of Physics and Astronomy \& LENS, University of Florence,} via G. Sansone 1, I-50019 Sesto Fiorentino, Italy.}

\begin{abstract}
The relation between the thermodynamic entropy production and non-Markovian evolutions is matter of current research. Here, we study the behavior of the stochastic entropy production in open quantum systems undergoing unital non-Markovian dynamics. In particular, for the family of Pauli channels we show that in some specific time intervals both the average entropy production and the variance can decrease, provided that the quantum dynamics fails to be P-divisible. Although the dynamics of the system is overall irreversible, our result may be interpreted as a transient tendency towards reversibility, described as a delta peaked distribution of entropy production around zero. Finally, we also provide analytical bounds on the parameters in the generator giving rise to the quantum system dynamics, so as to ensure irreversibility mitigation of the corresponding non-Markovian evolution.
\begin{description}
\item[PACS numbers] 03.65.Yz, 05.70.Ln, 42.50.Lc
\end{description}
\end{abstract}

\date{\today}

\maketitle

\section{Introduction}

In out-of-equilibrium settings the \textit{entropy production} is a fundamental thermodynamic quantity allowing to measure the degree of \textit{irreversibility} of the dynamical evolution of physical systems. Classically, in the framework of stochastic thermodynamics \cite{SeifertReview}, this irreversible contribution lies in the ratio (different from one) between the probability to observe a specific trajectory of the system and the probability to observe its time-reversed partner. The discrepancy between them is a consequence of an irreversible loss of the system internal energy, usually in the form of heat\,\cite{deGroot61,PRL2012}.

A similar framework of stochastic thermodynamics has been developed for quantum systems too\,\cite{NJP2010,PRL2011,SagawaLectures2014,PRL2015,FunoNJP2015,PRL2017,Gherardini_entropy,ManzanoPRX,PRL2018,Irreversibility_chapter,KimPRX2019}, where the trajectory has to be mapped into the sequence of outcomes of measurements performed on the system. This scheme, where just two measurements are taken into account, has been successfully used to prove a number of fluctuation relations that hold far from equilibrium and allow for the derivation of all the statistical moments of thermodynamic quantities\,\cite{EspositoReview,CampisiReview,2013Albash,2014Rastegin,NJP2015,Mehdi1,Mehdi2,2018GheraPRE,2019GomezArXiv,Cimini2020Arxiv}. In such a framework, relevant information about the dynamics of closed and open quantum systems can be extracted by looking at the probability distribution of the quantum entropy production\,\cite{HorowitzNJP2013,ManzanoPRE2015,Gherardini_entropy,ManzanoPRX}. In this respect, a dynamics is said \emph{reversible} if the latter distribution is a Dirac-delta in zero. This is the case if the system evolves according to a unitary dynamics and the effect due to possible measurements is negligible. When the dynamics is non-unitary and there are memory noise effects, one usually deals with non-Markovian quantum dynamics, which has become a topic of extensive research in the last decades\,\cite{RivasReview,BreuerReview,deVegaReview}. Apart from the theoretical foundational interest, this is also due to a number of experimental platforms where non-Markovianity turns out to be necessary to fully catch the relevant physics\,\cite{Exper1,LiuSciRep2013,Exper2,HaasePRL2018}.

Here, we address the relation between the non-Markovianity of the evolution of a quantum system and its thermodynamic reversibility, as described by the first two moments of the entropy production distribution. First results on this topic have started to appear quite recently\,\cite{BylickaSciRep2016,pezzutto,negentroSte,bhattacharya,ThomasPRE2018,popovic,CampbellPRA2018,strasberg,RivasThermo}, only considering the average entropy production rate. However, it is reasonable to expect that the non-Markovian character of the quantum dynamics may display some relevant features on the whole distribution of the entropy production, and not only on the $1$st moment. In this respect, we show that it is possible to have time intervals where both the $1$-th and $2$-nd cumulant of the quantum entropy distribution are decreasing, if the dynamics display a strong form of non-Markovianity known as \textit{essential non-Markovianity}\,\cite{degree}. 
It is worth noting that in this work we specifically concentrate on the case of unital quantum dynamics, which are customarily used to model open quantum systems subjected to white noise. This choice is motivated by the fact that the maximally mixed state (multiple of the identity) is a fixed point of the dynamics and represents a situation where the experimenter can access all the possible outcomes of a certain observable (e.g.\,of the energy) with equal probability. Moreover, on the thermodynamic side, these dynamics can be thought of describing the interaction of a quantum system with a thermal bath in the large (infinite) temperature limit.

The importance of our result lies in the following consideration: despite essentially non-Markovian evolutions allow for the existence of a time interval in which the average entropy production rate is negative\,\cite{negentroSte}, this does not necessarily imply a mitigation of irreversibility in general. Indeed, as it will be discussed below, there exist dynamical regimes in which, although the mean value of the entropy distribution decreases in a given time interval, the variance does not have the same behaviour. This implies that one can nevertheless face with high values of the entropy production on a single realization, occurring with low-probability but far from the corresponding average value that is decreasing. On the other hand, we also show that, already at the level of qubits, it is possible to have a transient reduction of both the average entropy production and the variance, thus signalling a tendency towards reversibility. An example and analytical bounds are provided to corroborate our analysis.

\section{Essential Non-Markovianity}
\label{sec:NM}

Many different approaches to quantum non-Markovianity can be found in the literature\,\cite{RivasReview, BreuerReview, deVegaReview}. In the following, we adopt the point of view first presented in Ref.\,\cite{RHP2010}, associating the concept of quantum Markovianity to the divisibility of the dynamical evolution. In particular, given a quantum evolution described by a one-parameter family of completely positive (CP) and trace preserving (TP) maps $\Lambda_t$, one says that the dynamics is CP-divisible if for any $t,s$ such that $t \geq s \geq 0$ one has $\Lambda_t = V_{t,s} \Lambda_s,$ with $V_{t,s}$ CP map. A dynamics is Markovian if it is CP-divisible, it is non-Markovian otherwise. Actually, one can go a step further and distinguish between different degrees of non-Markovianity, as suggested in\,\cite{degree}, depending on whether the intertwining map $V_{t,s}$ is $k$-positive, namely whether the map $V_{t,s}\otimes\mathrm{id}_k$ is positive ($\mathrm{id}_k$ is the identity map on $\mathbb{C}^k$) or not. Given a Hilbert space of dimension $n$, dynamical maps corresponding to $n$-positive $V_{t,s}$ are CP-divisible, those corresponding to $V_{t,s}$ that are only $1$-positive are called P-divisible, while if a dynamics is not even P-divisible we call it as \emph{essentially non-Markovian}. Resorting to a recently proved inequality\,\cite{PosMaps}, essential non-Markovianity is \textit{necessary} condition to find negative entropy production rates\,\cite{negentroSte}, even though it may not be sufficient\,\cite{popovic}. In this paper we consider unital dynamics, namely those evolutions that preserve the identity operator, fixed point of the map. Among them we focus on Pauli channels, whose Markovianity degree has been studied in detail in these recent papers\,\cite{PLA2013,PRA2015,PRA2016,BenattiPRA2017}.

\section{Non-Markovian Pauli channels}

In the following, we consider a two-level quantum system described by a density matrix $\varrho_t$ evolving in time through a unital dynamics. Notice that any unital qubit dynamics can be always described by a random unitary map\,\cite{unital2d,Caruso_RMP} (this is not true in higher dimension) that belongs to the family of Pauli channels. They are defined through the following Kraus representation:
\begin{equation}\label{map_Pauli_ch}
    \Lambda_{t}(\varrho) = \sum_{\alpha=0}^{3}p_{\alpha}(t)
    \sigma_{\alpha}\varrho\,\sigma_{\alpha},
\end{equation}
where $\{\sigma_{\alpha}\}_{0}^{3} = \{\mathbbm{1},\sigma_{x},\sigma_{y},\sigma_{z}\}$ is the set of Pauli matrices plus the identity, while the coefficients $p_\alpha$ obey the relation $\sum_{\alpha}p_{\alpha}(t)=1$, $\forall t$ (trace preservation). The initial condition $\Lambda_0= \mathrm{id}$ implies that $p_{0}(0)=1$ and $p_{\alpha}(0)=0$ for $\alpha\neq 0$. The map $\Lambda_{t}(\varrho)$ is CP if $p_{\alpha}(t)\geq 0$, $\forall t,\alpha$.

The conditions for CP-divisibility of the system dynamics are usually provided by introducing the generator $\mathcal{L}_{t}$ associated to the quantum map (\ref{map_Pauli_ch}). The generator satisfies the differential equation $\partial_t\Lambda_t = \mathcal{L}_{t}\Lambda_t$. Thus, under the hypothesis that the inverse $\Lambda_{t}^{-1}$ exists, the generator is defined as $\mathcal{L}_{t}=\partial_{t}\Lambda_{t}\circ\Lambda_{t}^{-1}$\,\cite{PLA2013} with $\circ$ denoting the composition of quantum maps. In this respect, one finds that the invertibility of the quantum map is ensured if $p_{1}+p_{2}$, $p_{2}+p_{3}$ and $p_{1}+p_{3}$ are different from $1/2$ for any finite time $t$. Moreover, being $p_{1}(0)=p_{2}(0)=p_{3}(0)=0$, the invertibility of the map together with the continuity in time of the functions $p_{\alpha}$ implies $p_{1}+p_{2}<1/2$, $p_{2}+p_{3}<1/2$ and $p_{1}+p_{3}<1/2$, $\forall t<\infty$. This working hypothesis will be assumed in the following. Note however that we may relax the invertibility assumption for asymptotically long times in order to produce a unique asymptotic state $\Lambda_{\infty}(\varrho)=\mathbbm{1}$, for any state $\varrho$.

As discussed in Appendix A, the Pauli channels generator can be written in the following general form\,\cite{PLA2013,degree,PRA2015}
\begin{equation}
    \mathcal{L}_{t}(\varrho) = \sum_{\alpha=1}^{3}\gamma_{\alpha}(t)
    \left(\sigma_{\alpha}\varrho\,\sigma_{\alpha}-\varrho\right),
\end{equation}
where $\gamma_{\alpha}$ are the so-called Lindblad coefficients. Therefore, the dynamics originated from $\Lambda_{t}(\varrho)$ is CP-divisible if and only if $\gamma_{\alpha}\geq 0$ $\forall t,\alpha$, while necessary and sufficient condition for P-divisibily is $\gamma_{\alpha}(t)+\gamma_{\beta}(t)\geq 0$ with $\alpha,\beta=1,2,3$ and $\alpha\neq\beta$. Finally, one also finds that the relations linking together all the Lindblad coefficients and the parameters $p_{\alpha}$ of the map are given by the following equation, with $\alpha,\beta=1,2,3$ and $\alpha\neq\beta$ (see also Appendix A):
\begin{equation}\label{relations_p_gamma}
    \exp\left(-2\int_{0}^{t}\gamma_{\alpha}(s)+\gamma_{\beta}(s)\,ds\right)
    = 1 - 2(p_{\alpha}(t)+p_{\beta}(t)).
\end{equation}

\section{Stochastic quantum entropy production}
\label{sec:entro}

The distribution of the quantum entropy production, originated by a generic quantum dynamics, can be obtained by realizing two distinct experimental procedures, i.e., a forward and a backward protocol that are appropriately chosen\,\cite{Gherardini_entropy,ManzanoPRX,Irreversibility_chapter,KimPRX2019}. Both protocols are interspersed by the application of two projective measurements at the initial and final time instants, according to the the well-known \textit{two-point measurement} (TPM) scheme\,\cite{CampisiReview}. The two measurements are defined as projections on the eigenstates of the arbitrary observables $\mathcal{O}_{\rm in}$ and $\mathcal{O}_{\rm fin}$. By using the spectral decomposition theorem, the observables $\mathcal{O}_{\rm in}$ and $\mathcal{O}_{\rm fin}$ can be generally written as $\mathcal{O}_{\rm in} = \sum_{k}a^{\rm in}_{k}\Pi^{\rm in}_{k}$ and $\mathcal{O}_{\rm fin} = \sum_{m}a^{\rm fin}_{m}\Pi^{\rm fin}_{m}$, where $\{\Pi\}$ is the set of projectors associated to the set of observable eigenvalues $\{a\}$ (measurement outcomes). The stochastic quantum entropy production $\Delta\sigma$ is then defined as\,\cite{Gherardini_entropy}:
\begin{equation}\label{general_sigma}
\Delta\sigma(a^{\textrm{fin}}_{m},a^{\textrm{in}}_{k}) \equiv \ln\frac{p_F(a^{\textrm{fin}}_{m},a^{\textrm{in}}_{k})}{p_B(a^{\textrm{in}}_{k},a^{\textrm{ref}}_{m})},
\end{equation}
where $p_F(a^{\textrm{fin}}_{m},a^{\textrm{in}}_{k})$ and $p_B(a^{\textrm{in}}_{k},a^{\textrm{ref}}_{m})$ are the joint probability to simultaneously measure the outcomes $\{a\}$ in a single realization of the forward and backward processes, respectively\,\cite{footnote1}. In Eq.\,(\ref{general_sigma}) $a^{\textrm{ref}}_{m}$ is obtained from the state after the $1$st measurement of the backward process, which is generally named as \textit{reference state}. Explicitly, the joint probabilities read
\begin{equation}
p_F(a^{\textrm{fin}}_{m},a^{\textrm{in}}_{k}) =
{\rm Tr}[\Pi^{\rm fin}_{m}\Lambda^F_{t_{\rm fin}}(\Pi^{\rm in}_{k})]p(a^{\rm in}_{k})
\end{equation}
and
\begin{equation}
p_B(a^{\textrm{in}}_{k},a^{\textrm{ref}}_{m}) =
{\rm Tr}[\Pi^{\rm in}_{k}\Lambda^B_{t_{\rm fin}}(\Pi^{\rm fin}_{m})]p(a^{\rm ref}_{m}).
\end{equation}
If the CPTP map $\Lambda_t$ governing the system dynamics is unital, then it is customary to consider the backward dynamics $\Lambda_t^B$ as the dual of the forward one (because it is itself a proper quantum dynamics). As a result, the stochastic quantum entropy production $\Delta\sigma$ becomes $\Delta\sigma(a^{\textrm{fin}}_{m},a^{\textrm{in}}_{k}) = \ln\left(p(a^{\textrm{in}}_{k})/p(a^{\textrm{ref}}_{m})\right)$, with $p(a^{\textrm{ref}}_{m})$ denoting the probability to get the measurement outcome $a_{m}^{\rm ref}$. It is reasonable to choose the reference state equal to the final density operator after the $2$nd measurement of the forward process. This means that for our purposes the stochastic quantum entropy production is equal to
\begin{equation}\label{stochentro}
\Delta\sigma(a^{\textrm{fin}}_{m},a^{\textrm{in}}_{k}) = \ln p(a^{\textrm{in}}_{k}) - \ln p(a^{\textrm{fin}}_{m}),
\end{equation}
where $p(a^{\textrm{fin}}_{m})$ denotes the probability to measure the $m-$th outcome at the final time instant $t_{\rm fin}$.

The statistics of the stochastic quantum entropy production $\Delta\sigma$ can be computed by evaluating the corresponding probability distribution $\textrm{Prob}(\Delta\sigma)$. Each time we repeat the TPM scheme, one has a different realization for $\Delta\sigma$ within a set of discrete values, whereby $\textrm{Prob}(\Delta\sigma)$ is fully determined by the knowledge of the measurement outcomes and the respective probabilities:
\begin{equation}\label{prob_sigma}
\textrm{Prob}(\Delta\sigma) = \sum_{k,m}\delta\left[\Delta\sigma - \Delta\sigma(a^{\textrm{fin}}_{m},a^{\textrm{in}}_{k})\right]p(a_{k}^{\rm in},a_{m}^{\rm fin}),
\end{equation}
where $\delta[\cdot]$ denotes the Dirac delta and $p(a_{k}^{\rm in},a_{m}^{\rm fin}) =
{\rm Tr}[\Pi^{\rm fin}_{m}\Lambda_{t_{\rm fin}}(\Pi^{\rm in}_{k})]p(a^{\rm in}_{k})$ with $p(a^{\rm in}_{k}) = {\rm Tr}[\varrho_{0}\Pi^{\rm in}_{k}]$.

All the statistical moments of $\Delta\sigma$ can be obtained by using the characteristic function $G_{\Delta\sigma}(u) \equiv \int\textrm{Prob}(\Delta\sigma)e^{iu\Delta\sigma}{\rm d}\Delta\sigma$ with $u\in\mathbb{C}$. As formally shown in Appendix B, there exists a \textit{closed-form expression} for each quantum entropy statistical moment, provided that a TPM scheme is used to derive the entropy fluctuations. As a consequence, one can determine the $1$st and $2$nd moments of $\Delta\sigma$. The former equals to
\begin{eqnarray}\label{eq_first_moment}
\langle\Delta\sigma\rangle &=& -{\rm Tr}[\ln\varrho_{\tau}\Lambda_{t_{\rm fin}}(\varrho_{\rm in})] + {\rm Tr}[\varrho_{\rm in}\ln\varrho_{\rm in}]\nonumber \\
&=& \Delta S + S(\varrho_{\rm fin}\|\varrho_{\tau})
\end{eqnarray}
with $S(\varrho\|\sigma)$ denoting the quantum relative entropy of $\varrho$ with respect to $\sigma$ and $\Delta S \equiv S(\varrho_{\rm fin}) - S(\varrho_{\rm in})$ the difference of the von-Neumann entropies of $\varrho_{\rm in}$ and $\varrho_{\rm fin}$. In Eq.\,(\ref{eq_first_moment}),
\begin{equation*}
\varrho_{\rm in} = \sum_{k}p(a_{k}^{\rm in})\Pi_{k}^{\rm in}\,\,\,\,\,\text{and}\,\,\,\,\,\varrho_{\tau} = \sum_{m}p(a_{m}^{\rm fin})\Pi_{m}^{\rm fin}
\end{equation*}
are, respectively, the ensemble average of the quantum system after the $1$st and $2$nd measurements of the TPM scheme, while $\varrho_{\rm fin} \equiv \Lambda_{t}(\varrho_{\rm in})$ is the density operator before the $2$nd projective measurement. Instead, the $2$nd statistical moment of $\Delta\sigma$ is given by the following relation:
\begin{eqnarray}
&\langle\Delta\sigma^2\rangle = {\rm Tr}[(\ln\varrho_{\tau})^2\Lambda_{t_{\rm fin}}(\varrho_{\rm in})]&\nonumber \\
&-2{\rm Tr}[\ln\varrho_{\tau}\Lambda_{t_{\rm fin}}(\varrho_{\rm in}\ln\varrho_{\rm in})]+{\rm Tr}[\varrho_{\rm in}(\ln\varrho_{\rm in})^2].&
\end{eqnarray}
In this manuscript, we will mostly focus on the variance that is related to the $2$nd moment as usual, i.e.,
\begin{equation}
    {\rm Var}(\Delta \sigma) \equiv \langle\Delta\sigma^2\rangle - \langle\Delta\sigma\rangle^2.
\end{equation}

\section{Pauli channels and stochastic entropy}

We apply the formalism of the stochastic thermodynamics to the Pauli channel model. For the sake of convenience, we take the observable $\mathcal{O}$, associated to both the quantum projective measurements of the TPM, equal to the Pauli operator $\sigma_z$. As a result, the projectors $\Pi^{\rm in}$ and $\Pi^{\rm fin}$ are described by the pure states $|\ell\rangle\!\langle \ell|$ with $\ell\in\{0,1\}$, whereby
\begin{equation*}
|0\rangle\!\langle 0|=\frac{\mathbbm{1}+\sigma_{z}}{2}\,\,\,\,\,\text{and}\,\,\,\,\,|1\rangle\!\langle 1|=\frac{\mathbbm{1}-\sigma_{z}}{2}\,.
\end{equation*}
By initializing the system in the state $\varrho_0$, the $1$st measurement of the TPM scheme makes the quantum system collapse in one of the eigenstates of $\sigma_{z}$. Thus, the ensemble average of the system after such a measurement is given by the mixed state $\varrho_{\rm in}$ with diagonal elements $\frac{1+\zeta_0}{2}$ and $\frac{1-\zeta_0}{2}$, where $\zeta_0 \equiv1 - 2\varrho_{0}^{(11)} $ and $\varrho_{0}^{(11)} \equiv \langle 1| \varrho_{0} |1\rangle$. Since
\begin{equation*}
\Lambda_{t}(|0\rangle\!\langle 0|)=\frac{\mathbbm{1}+2(1/2-p_{1}-p_{2})\sigma_{z}}{2}
\end{equation*}
with $p_1$, $p_2$ coefficients defining a quantum Pauli channel, also $\varrho_{\rm fin}=\Lambda_{t}(\varrho_{\rm in})$ is a mixed state $\forall t$, with $\frac{1+\zeta_t}{2}$ and $\frac{1-\zeta_t}{2}$ being its diagonal elements. Here, $\zeta_t \equiv \lambda_{t}\zeta_0$ where
\begin{equation}
    \lambda_t \equiv 1-2(p_{1}(t)+p_{2}(t))
    =\mathrm{e}^{-2\int_0^t (\gamma_1(s) + \gamma_2(s))\mathrm{d}s}
\end{equation}
with $0 \leq \lambda_t \leq 1$. The upper bound is saturated at time $t=0$, while the lower bound is never achieved at finite time but it turns out that $\lambda_t$ converges to $0$ for $t \to \infty$, thus implying that $\lambda_{\infty}=0$.

Moreover, being $[\sigma_z,\varrho_{\rm fin}]=0$, one has $\varrho_{\tau}=\varrho_{\rm fin}$. Hence, simplifying the expressions of $\langle\Delta\sigma\rangle$ and $\langle\Delta\sigma^2\rangle$, we obtain $\langle\Delta\sigma\rangle = \Delta S$, while the $2$nd moment of $\Delta\sigma$, in case of initial pure state with $\zeta_0=1$, reads
\begin{equation}
\langle \Delta \sigma^2 \rangle = \mathrm{Tr}\left[ (\ln\varrho_{\rm fin})^{2}\Lambda_t(\varrho_{\rm in}) \right].
\end{equation}
As final remark, it is worth noting that, by fixing the initial state ($\zeta_0=1$) to be pure, there is an apparent asymmetry between the forward and backward processes that could entail realizations of the stochastic entropy production with values tending to infinite. However, the probability that these realizations can occur is vanishing. This evidence is crucial in explaining the convergence of our results and the admissibility of our choice. Indeed, the formalism enabling the computation of the mean and variance of the stochastic entropy production adopts the convention $0\ln 0 := 0$, which prevents entropy divergent behaviours.   

\section{Mitigating thermodynamic irreversibility}
\label{sec:result}

In this paragraph, we present our results relating non-Markovianity and stochastic entropy production. Depending on the final time $t$ of the TPM scheme, one has a different probability distribution for the entropy production, with an average value that typically increases with time (this always occurs if the dynamics is P-divisible) and a non-monotonic behaviour on the variance. Below, we are going to show that it is possible to have a time interval in which both the average and the width of the distribution decrease, thus signaling an irreversibility mitigation. 
In particular, we will compute explicitly the time-derivative of the first two statistical cumulants $\langle\Delta\sigma\rangle$ and ${\rm Var}(\Delta \sigma)$, and then evaluate their sign.

As first step, we start from the computation of the time-derivative of $\langle\Delta\sigma\rangle$. The latter, for the class of models we consider, equals to
\begin{eqnarray}
&\displaystyle{\Delta S \equiv S(\varrho_{\rm fin}) - S(\varrho_{\rm in})=}&\nonumber \\
&\displaystyle{=\frac{1+\zeta_0}{2}\ln\left(\frac{1+\zeta_0}{2}\right)+
\frac{1-\zeta_0}{2}\ln\left(\frac{1-\zeta_0}{2}\right)}&\nonumber \\
&\displaystyle{-\frac{1+\zeta_t}{2}\ln\left(\frac{1+\zeta_t}{2}\right)-\frac{1-\zeta_t}{2}\ln\left(\frac{1-\zeta_t}{2}\right)}&
\end{eqnarray}
where $\zeta_0 $ and $\zeta_t$ have been defined in the previous section.
For simplicity, let us assume $\zeta_0=1$. As we will show below, this simple choice of the initial state is just sufficient to find an evidence of the mechanism ruling irreversibility mitigation. Notice that for a non-P divisible unital quantum map the time-derivative of $\langle\Delta\sigma\rangle$ is not necessarily positive. Indeed, $\partial_t \langle\Delta\sigma\rangle$ explicitly reads
\begin{equation}
\partial_t \langle\Delta\sigma\rangle = \lambda_t\ln\left(\frac{1+\lambda_t}{1-\lambda_t}\right)(\gamma_1(t)+\gamma_2(t)) 
\end{equation}
with $\lambda_t\ln(\frac{1+\lambda_t}{1-\lambda_t})$ always non negative, so that $\partial_t \langle\Delta\sigma\rangle$ is negative whenever the sum $\gamma_1(t)+ \gamma_2(t)$ becomes negative. This happens when the dynamics fails to be P-divisible. Let us observe that, while the time-derivative of the average entropy production is divergent for $\lambda_t=1$ (occurring only at $t=0$), $\langle\Delta\sigma\rangle$ always takes finite values, again due to the properties of the function $x\ln x$. 

As a second step, in order to see some effects on the system reversibility due to non-Markovianity, we look at the time-derivative of the variance ${\rm Var}(\Delta \sigma)$ and study its sign. By substituting the expressions of $\varrho_{\rm in}$ and $\varrho_{\rm fin}$ (depending on $\lambda_t$) in the formula for the second moment $\langle\Delta\sigma^2\rangle = \mathrm{Tr}\left[ (\ln\varrho_{\rm fin})^{2}\Lambda_t(\varrho_{\rm in}) \right]$, one finds that
 \begin{equation}
    \partial_t  \langle \Delta \sigma^2 \rangle = -2  \left(1+\frac{1}{2}\ln\Big(\frac{1-\lambda_t^2}{4} \Big) \right) \partial_t \langle \Delta \sigma \rangle \,.
 \end{equation}
As a consequence, the time-derivative of the stochastic entropy variance reads
\begin{equation}\label{app_variance_deriv}
  \partial_t {\rm Var}(\Delta \sigma) = 2 f_{t} \,\partial_t \langle \Delta \sigma \rangle \,,
\end{equation}
where the function $f_{t}$ is defined as follows
\begin{align}
  f_{t} &\equiv - \left(  \langle\Delta\sigma\rangle+1+\frac{1}{2}\ln\Big(\frac{1-\lambda_t^2}{4} \Big)\right)  \nonumber \\
  & =\frac{\lambda_t}{2}\ln{\left(\frac{1+\lambda_t}{1-\lambda_t}\right)}-1 \,.
\end{align}
Thus, given the sign of $\partial_t \langle \Delta \sigma \rangle$, one can also determine the sign of the variance by looking at the function $f_{t}$. On the one hand, the function $f_t$ is known to be always greater or equal than $-1$ and is such that $\lim_{t \to 0} f_t= +\infty$ and $\lim_{t \to \infty} f_t = -1$. On the other hand, by computing $\partial_t f_t$\,, namely
 \begin{equation*}\label{SM_eq:derivative_f}
    \partial_t f_{t} =  \left( \frac{1}{2}\ln{\left(\frac{1+\lambda_t}{1-\lambda_t}\right)}  + \frac{\lambda_t}{1-\lambda_t^2} \right)\partial_t{\lambda_t}\,,
\end{equation*}
one observes that $f_t$ is increasing or decreasing depending of $\partial_t{\lambda}_t$. In particular, it is increasing in the region where $\gamma_1(t)+ \gamma_2(t)$ is negative that means when P-divisibility is broken. Therefore, assuming $\gamma_1 + \gamma_2 \leq 0$ in a single interval $[t_1,t_2]$, the function $f_t$ is decreasing up to time $t_1$, then increases from $t_1$ to $t_2$ and finally decreases for $t>t_2$. As a result, one can have three different cases for the sign of $f_t$\,:
\begin{eqnarray}
    &{\rm (I)}\,\,\,f_{t_1}\geq 0\,\,\,\text{and}\,\,\,f_{t_2}\geq 0\,;\,\,\,\,\,\,\,
    {\rm (II)}\,\,\,f_{t_1} < 0\,\,\,\text{and}\,\,\,f_{t_2}\geq 0\,;&\nonumber \\
    &{\rm (III)}\,\,\,f_{t_1} < 0\,\,\,\text{and}\,\,\,f_{t_2} < 0\,.&
\end{eqnarray}
\begin{figure}[t!]
\begin{center}
\includegraphics[scale=0.475]{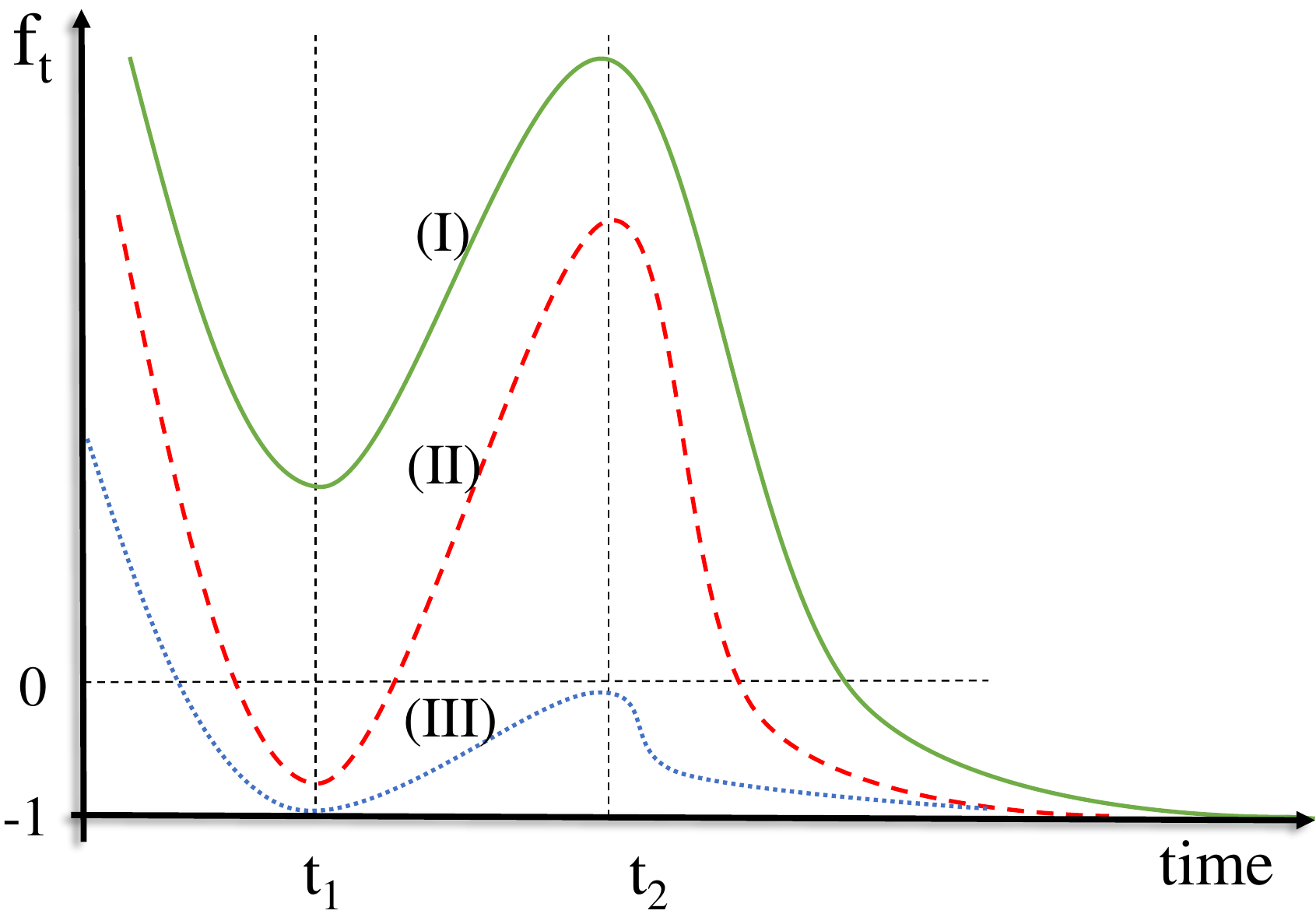}
\caption{Pictorial representation of the three cases {\rm (I)}, {\rm (II)} and {\rm (III)}, respectively green solid, red dashed and blue dotted lines, concerning the possible behaviour of $f_t$ as a function of time $t$.}
\label{fig:figure1}
\end{center}
\end{figure}
In Fig.\,\ref{fig:figure1} we report a sketch of the possible behaviour of $f_t$ as a function of time in the three different cases {\rm (I)}, {\rm (II)} and {\rm (III)}, corresponding to three different situations for the sign of $\partial_t {\rm Var}(\Delta \sigma)$ in the interval $[t_1,t_2]$, where the time-derivative of the average is also negative. In particular, one has:
\begin{eqnarray}\label{derivative_variance}
    &&{\rm (I)}\,\,\,\partial_t {\rm Var}(\Delta \sigma)\leq 0\,\,\,\text{in}\,\,\,[t_1,t_2]\nonumber \\
    &&{\rm (II)}\,\,\,\partial_t {\rm Var}(\Delta \sigma)\leq 0\,\,\,\text{in}\,\,\,[t_3,t_2]\,\,\,\text{with}\,\,\,t_1<t_3<t_2 \nonumber \\
    &&{\rm (III)}\,\,\,\partial_t {\rm Var}(\Delta \sigma)\geq 0\,\,\,\text{in}\,\,\,[t_1,t_2].
\end{eqnarray}
In cases (I) and (II) there is a time interval in which the system tends to be more reversible, in the sense that both the average and the variance of $\textrm{Prob}(\Delta\sigma)$ are reducing, so that the distribution becomes sharper. As a matter of fact, the reversibility of a quantum system dynamics is associated to a shrinking of the quantum entropy distribution $\textrm{Prob}(\Delta\sigma)$ up to approach a Dirac delta $\delta\left[\Delta\sigma\right]$. So, the decreasing of $\partial_t\langle\Delta\sigma\rangle$ and $\partial_t {\rm Var}(\Delta\sigma)$ in a given time interval represents an evident tendency towards reversibility in the transient, induced by the presence of non-Markovian effects.

Now, we provide analytical bounds on the coefficients $p_{\alpha}(t)$ of the Pauli channel that are sufficient to mitigate irreversibility. Above, provided that the dynamics of the system is not P-divisible, we have shown that the tendency of ${\rm Var}(\Delta \sigma)$ to decrease just depends on the sign of $f_t$. By introducing $\phi_t \equiv \int_{0}^{t}(\gamma_1(s) + \gamma_2(s))\mathrm{d}s$, the inequality $f_t \geq 0$ can be recast in the relation
\begin{equation}\label{ineq_varphi}
e^{-2\phi_t}\ln\left(\frac{1+e^{-2\phi_t}}{1-e^{-2\phi_t}}\right) \geq 2 \,.
\end{equation}
The function $x\ln(\frac{1+x}{1-x})-2 $, with $0 \leq x < 1$, has an unique zero at $x^{\ast} \approx 0.8336$ and is positive for $x \geq x^{\ast}$. This implies that the inequality (\ref{ineq_varphi}) is verified for $x \geq x^{\ast}$, i.e.,
\begin{equation}\label{ineq_varphi_2}
    0 \leq \phi_t \leq \phi^{\ast} \equiv - \frac{1}{2}\ln(x^{\ast}) \approx 0.091
\end{equation}
for all $t>0$. Eq.\,(\ref{ineq_varphi_2}) clearly shows that the irreversibility mitigation can be found only in a quite small range of dynamical parameters. This means that {\it essential non-Markovianity} has to be usually associated to irreversibility, except some narrow regimes whereby a transient tendency to reversibility could be observed.

\section{Analytical example}

Here, we present an example of legitimate (namely completely positive and trace preserving) unital dynamics for a qubit such that the evolution is not P-divisible in a single time interval $[t_1,t_2]$. As discussed before, one has to satisfy the following constraints:
\begin{eqnarray}
&&{\rm (i)}\,\,\,\gamma(t)= \gamma_1(t)+\gamma_2(t) \leq 0\,\,\,\text{in}\,\,\,[t_1,t_2]\,\,\,\text{(no P-divisibility)}\nonumber \\
&&{\rm (ii)}\,\,\,\phi_t \geq 0\,\,\,\text{for any}\,\,\,t\geq 0\,\,\,\text{(CP dynamics)}.\nonumber
\end{eqnarray}
This in turn implies that $\lambda_t = \mathrm{e}^{-2\phi_t} \leq 1$. We assume the following explicit form for the function $\gamma(t)$
\begin{equation}\label{expl_form_gamma}
\gamma(t)= \beta - \mathrm{e}^{-\alpha t}\left(1- \mathrm{e}^{-\alpha t} \right),
\end{equation}
where $\alpha,\beta > 0$ are two positive parameters. As a consequence the function $\phi_t$ reads 
\begin{equation*}
\phi_t = \beta t - \frac{\left( 1- \mathrm{e}^{-\alpha t} \right)^2}{2\alpha}\,.
\end{equation*}
Then, the sign of $\gamma(t)$ can be easily studied. In particular, one finds that two zeros exist at times $t_1$ and $t_2$ corresponding to

\begin{small}
\begin{equation}
t_1 = -\frac{1}{\alpha} \ln\left( \frac{1}{2} + \frac{\sqrt{1-4\beta}}{2} \right), \quad t_2 = -\frac{1}{\alpha} \ln\left( \frac{1}{2} - \frac{\sqrt{1-4\beta}}{2} \right)
\end{equation}
\end{small}

provided that $\beta < \frac{1}{4}$. Moreover, it turns out that $\gamma(t)$ is negative between $t_1$ and $t_2$ and positive otherwise, thus satisfying condition {\rm (i)}. Instead, condition {\rm (ii)} corresponds to the requirement $\beta \geq \frac{\left( 1- \mathrm{e}^{-\alpha t} \right)^2}{2\alpha t}$, $\forall t>0$. Therefore, one has to impose a lower bound $\overline{\beta}$ to $\beta$, which is given by
\begin{equation}
\overline{\beta} = \max_{t>0} \frac{\left( 1- \mathrm{e}^{-\alpha t} \right)^2}{2\alpha t}.
\end{equation}
\begin{figure}[t!]
\begin{center}
\includegraphics[scale=0.63]{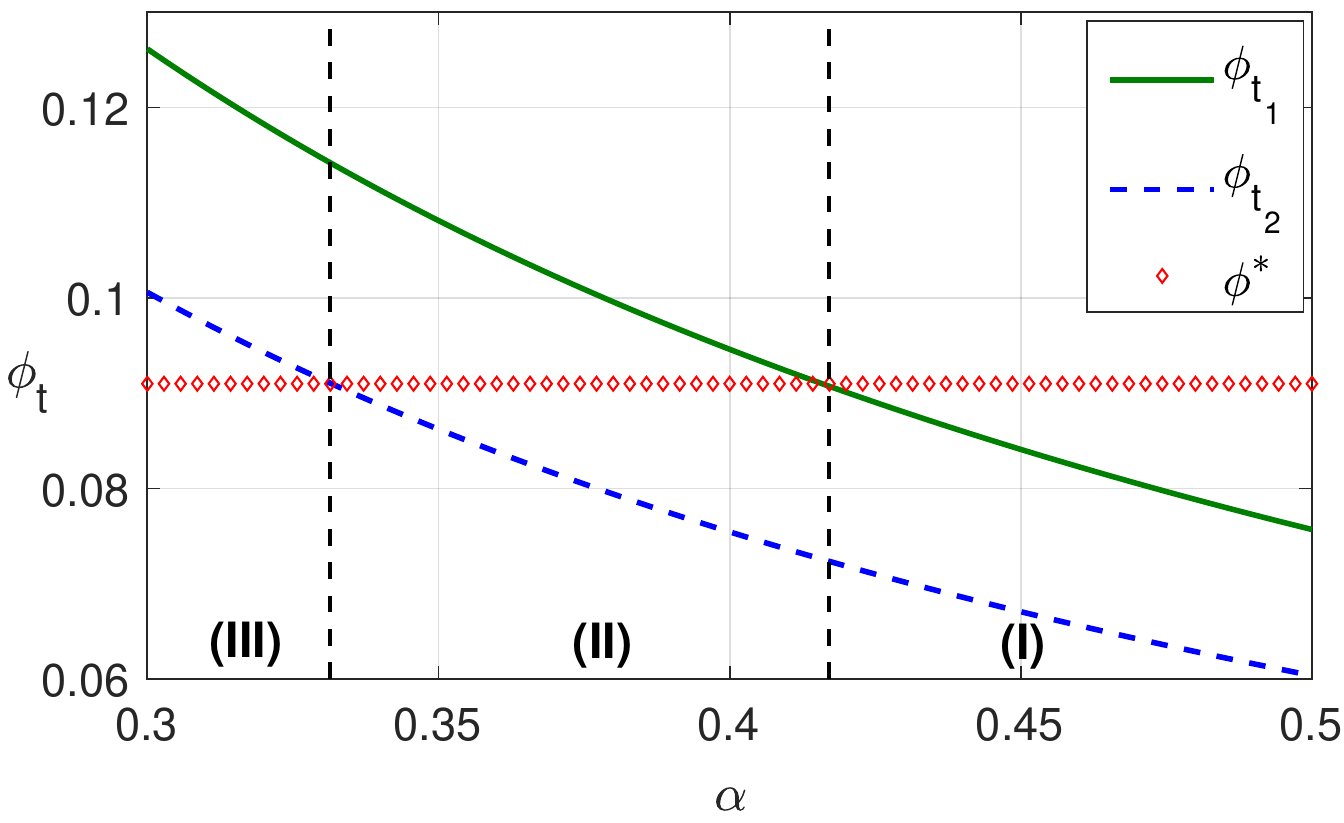}
\caption{Plot of the values of $\phi_{t_{1}}$ (green solid line) and $\phi_{t_2}$ (blue dashed line) as a function of $\alpha$, with $\beta=0.23$ ($0.2 <\beta< 0.25$). The horizontal line (red diamonds) corresponds to $\phi^{\ast}$ and allows to distinguish the three different cases {\rm (I)}, {\rm (II)} and {\rm (III)} for the sign of $\partial_t {\rm Var}(\Delta \sigma)$.}
\label{fig:figure2}
\end{center}
\end{figure}
It can be easily found that the maximum of the function is implicitly defined by the relation $\mathrm{e}^{z_{\rm max}}=1+2z_{\rm max}$, with $z_{\rm max} \equiv \alpha t_{\rm max}$. Numerically, one obtains the value $z_{\rm max} \approx 1.25$ that in turn implies $\overline{\beta} = 2\,z_{\rm max}\,\mathrm{e}^{-2z_{\rm max}} \approx 0.2$ for any value of $\alpha$. As a result, conditions {\rm (i)} and {\rm (ii)} bound the parameter $\beta$ to be
\begin{equation}
\overline{\beta} < \beta < \frac{1}{4}
\end{equation}
because $\overline{\beta}\approx 0.2$ is a nontrivial lower bound smaller than $1/4$. 
\begin{figure}[t!]
\begin{center}
\includegraphics[scale=0.54]{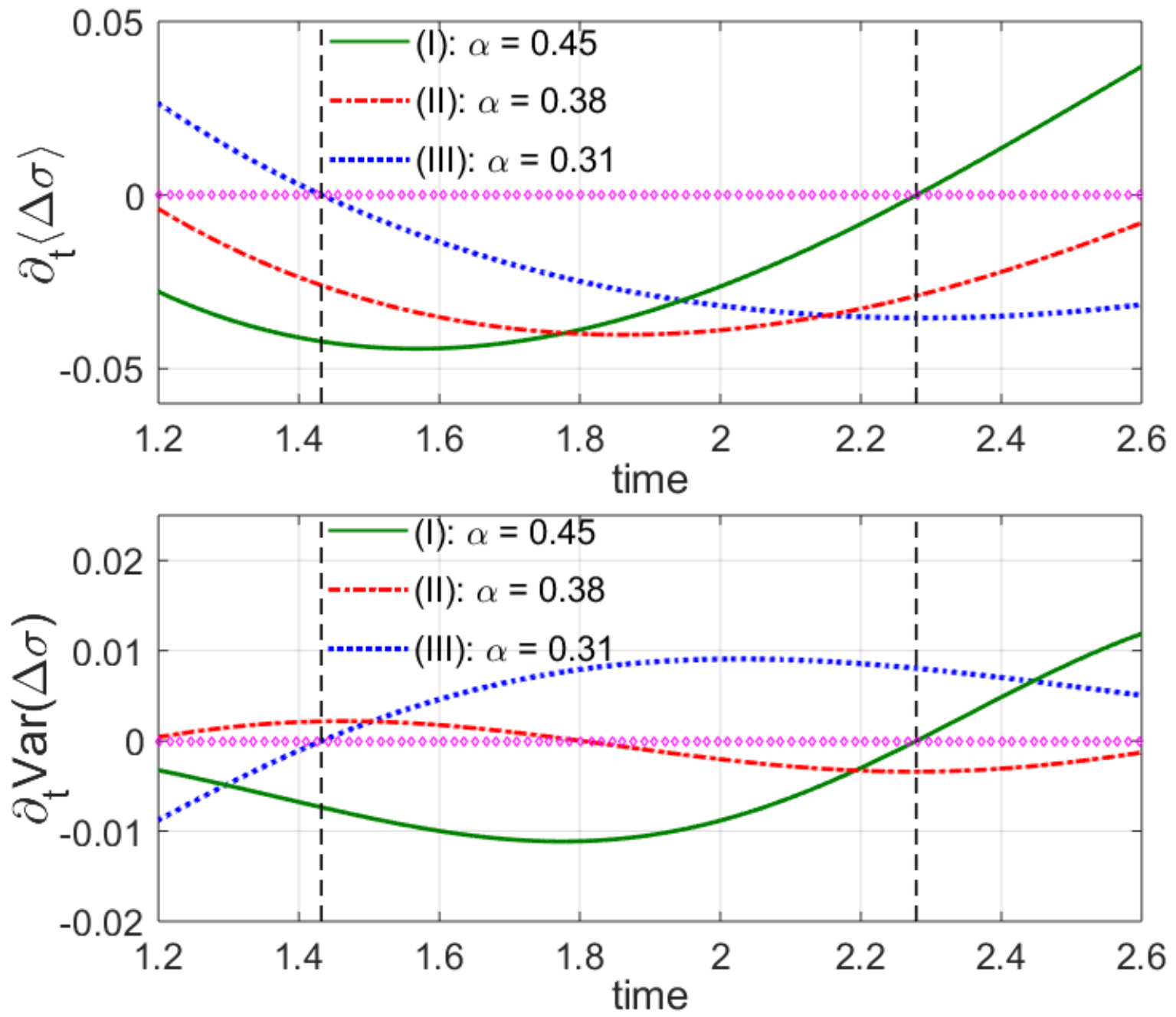}
\caption{Time-derivatives of $\langle\Delta\sigma\rangle$ and ${\rm Var}(\Delta\sigma)$, respectively average entropy production and entropy variance, for the analytical example of Section VII. Given the explicit form of the function $\gamma(t)$ (Eq.\,(\ref{expl_form_gamma})), we take $\beta=0.23$ as in Fig.\,\ref{fig:figure2} and three values of $\alpha$, i.e.\,$0.31,0.38,0.45$, corresponding to the cases (I), (II) and (III), respectively represented by green solid, red dash-dotted and blue dotted lines. In each of them, the time-derivative of the entropy variance has a different behaviour in the time interval $[1.43,2.28]$ where $\partial_{t}\langle\Delta\sigma\rangle$ is negative.}
\label{fig:figure3}
\end{center}
\end{figure}
As shown in Fig.\,\ref{fig:figure2}, one can span the three different regimes {\rm (I)}, {\rm (II)} and {\rm (III)} for the sign of the variance time-derivative, by tuning the parameter $\alpha$. According to Eq.\,(\ref{ineq_varphi_2}), these regimes are obtained by comparing the values $\phi_{t_{1}}(\alpha)$ and $\phi_{t_{2}}(\alpha)$ with $\phi^{\ast} = -\frac{1}{2}\ln(x^*) \approx 0.091$, which is the value corresponding to a vanishing function $f_t$. Finally, to corroborate the results of our analysis, in Fig.\,\ref{fig:figure3} we plot the time-derivatives of $\langle\Delta\sigma\rangle$ and ${\rm Var}(\Delta\sigma)$ as a function of time, for $\beta=0.23$ and $\alpha\in\{0.31,0.38,0.45\}$ corresponding, respectively, to the regimes (I), (II) and (III). As shown in the figure, there exist at least one time interval $[t_1,t_2]$ (in the example, $t_1=1.43$ and $t_2=2.28$) where the time-derivative of the average entropy production is negative for the three chosen values of $\alpha$. Nevertheless, the time-derivative of the entropy variance ${\rm Var}(\Delta\sigma)$ has not the same behaviour. Indeed, as predicted by our theoretical analysis, only cases (I) and (II) allow for negative values of $\partial_{t}{\rm Var}(\Delta\sigma)$ within $[t_1,t_2]$. Once again, this evidence shows that in a non-Markovian quantum dynamics the mitigation of the thermodynamic irreversibility (i) occurs only for specific values of the parameters governing the dynamics of the system, and (ii) cannot be just ensured by the negativity of the average entropy rate. 

\section{Conclusions}

We have further investigated on the relations between entropy production and non-Markovianity using the formalism of stochastic thermodynamics. We have shown that it is possible to have legitimate non-Markovian dynamics, namely 1-parameter families of completely positive and trace preserving maps that allow for both the average entropy production and its variance to be transiently decreasing. This can happen when the dynamics is not P-divisible. Being a dynamics reversible if the distribution of the entropy production is a Dirac delta, we interpret our finding as a transient tendency to reversibility. Our analysis deals with unital qubit dynamics, for which we provide analytical bounds in the parameter space corresponding to irreversibility mitigation. The calculation is done assuming a pure initial state because already in this simple case we find evidence of the phenomenon we are interested in. As a concluding remark, we also note that, provided that the system dynamics is not P-divisible, it could be in principle possible to find legitimate dynamics that allow for the same phenomenology both in non-unital dynamics and higher dimensional quantum systems. This will be matter for future investigation.

\begin{acknowledgements}
S.G. and S.M. equally contributed to this work. The authors thank F. Benatti for fruitful discussions and F. Carollo for useful comments on the manuscript.
S.G. and F.C. were financially supported by the Fondazione CR Firenze through the projects Q-BIOSCAN and QUANTUM-AI, PATHOS EU H2020 FET-OPEN Grant No.\,828946, the UNIFI Grant Q-CODYCES, and the MISTI Global Seed Funds MIT-FVG Collaboration Grant ``NVQJE''. S.M. was financially supported by the EPSRC Grant no.\,EP/R04421X/1.
\end{acknowledgements}

\appendix

\section*{Appendices}

\subsection*{A. Brief overview on Pauli channels}
\label{app_A}

Pauli channels are quantum dynamical maps of the form
\begin{equation}
    \Lambda_{t}(\varrho) = \sum_{\alpha=0}^{3}p_{\alpha}(t)
    \sigma_{\alpha}\varrho\,\sigma_{\alpha},
\end{equation}
where $\{\sigma_{\alpha}\}_{0}^{3} = \{\mathbbm{1},\sigma_{x},\sigma_{y},\sigma_{z}\}$ is the set of Pauli matrices plus the identity, the coefficients $p_\alpha$ obey the relation $\sum_{\alpha}p_{\alpha}(t)=1$, $\forall t$ (trace preservation), and the initial condition $\Lambda_{0}= {\rm id}$ enforces $p_0(0)=1$. Each map $\Lambda_{t}$ is completely positive if $p_\alpha (t) \geq 0 \, \forall \,\alpha, t\geq 0$. One can easily check that the Pauli matrices are the eigen-operators of the linear map $\Lambda_t$ and, in particular, one has
\begin{align}
    &\Lambda_t(\mathbbm{1})= \mathbbm{1}, \\
    &\Lambda_t(\sigma_{1})=(1-2p_2(t)-2p_3(t))\sigma_{1}, \\
    &\Lambda_t(\sigma_{2})=(1-2p_1(t)-2p_3(t))\sigma_{2}, \\
    &\Lambda_t(\sigma_{3})=(1-2p_1(t)-2p_2(t))\sigma_{3}.
\end{align}
The map is invertible provided that $p_1(t)+p_2(t)\neq 1/2,\, p_1(t)+p_3(t)\neq 1/2,\, p_2(t)+p_3(t)\neq 1/2 $ at any time $t>0$. Since initially the coefficients $p_i$ with $i\in\{1,2,3\}$ are vanishing (because $p_0(0)=1$) we can enforce continuity of the functions $p_i(t)$ and invertibility of the map $\Lambda_t$ at any time if the constraints $p_1(t)+p_2(t)<1/2,\, p_1(t)+p_3(t)< 1/2,\, p_2(t)+p_3(t)< 1/2 $ are satisfied. For an invertible dynamics the time-dependent generator turns out to be $\mathcal{L}_t= \partial_t \Lambda_t \circ \Lambda_t^{-1}$. By comparing the following ansatz for the generator
\begin{equation}
    \mathcal{L}_t(\varrho)= \sum_{i=1}^3 \gamma_i(t)\big( \sigma_i \varrho \sigma_i  - \varrho \big),
\end{equation}
with the expression derived computing $\partial_t \Lambda_t$ and $\Lambda^{-1}_t$ one obtains the following relation between the functions $\gamma_i(t)$ and the functions $p_i(t)$
\begin{equation}
    \gamma_i(t) + \gamma_j(t)= \frac{\partial_t(p_i(t) + p_j(t))}{1-2(p_i(t)+p_j(t))}
\end{equation}
for any pair $i,j$ with $i\neq j$. The previous differential equations can be easily integrated by recognizing that
\begin{equation}
    \frac{\partial_t(p_i(t) + p_j(t))}{1-2(p_i(t)+p_j(t))}= -\frac{1}{2}\partial_t\ln\big(1-2p_{i}(t)-2p_{j}(t)\big),
\end{equation}
so that finally one has
\begin{equation}
    p_i(t) + p_j(t) = \frac{1-\mathrm{e}^{-2\int_0^t (\gamma_i +\gamma_j)\mathrm{d}s}}{2}.
\end{equation}
Therefore, we have two equivalent characterizations of the dynamics, one based on the Lindblad coefficients $\gamma_i$ and the other one based on the parameters $p_i$ of the Krauss decomposition, and we know how to connect the two. This is important because the conditions for complete positivity (CP) are easily given for the $p_i$ while conditions for CP-divisibility and P-divisibility are given on the $\gamma_i$. These conditions are reported in the main text, as first derived in Ref.\,\cite{PRA2015}.

\subsection*{B. Closed-form expression of quantum entropy statistical moments}
\label{app_B}

The statistical moments of a random variable $X$ with probability distribution $\textrm{Prob}(X)$ can be generally computed by introducing the characteristic function
\begin{equation}
G_{X}(u) \equiv \int\textrm{Prob}(X)e^{iuX}{\rm d}X
\end{equation}
associated to $\textrm{Prob}(X)$, with $u$ complex number. In this regard, it holds that
$\langle X^{\ell}\rangle = \left.(-i)^{\ell}\partial_{u}^{\ell}G_{X}(u)\right|_{u=0}$
namely the $\ell$-th statistical moment of $X$ is proportional to the $\ell$th derivative of $G_{X}(u)$ with respect to $u$ and evaluated at $u=0$.
This property can be thus applied to the computation of the statistical moments of $\Delta\sigma$ so that we are allowed to write
\begin{equation}
\langle\Delta\sigma^{\ell}\rangle = \left.(-i)^{\ell}\partial_{u}^{\ell}G_{\Delta\sigma}(u)\right|_{u=0}.
\end{equation}
Provided that a TPM scheme is used to derive the fluctuations of entropy, here we show that there exists a \textit{closed-form expression} for each quantum entropy statistical moment. In particular, the $\ell$-th statistical moment $\langle\Delta\sigma^{\ell}\rangle$, with $\ell\geq1$ ($\ell$ arbitrary integer), is equal to
\begin{eqnarray}\label{moments}
&\langle\Delta\sigma^{\ell}\rangle =&\nonumber \\ &=\displaystyle{\sum_{n=0}^{\ell}(-1)^{\ell-n}\binom{\ell}{n}{\rm Tr}\left[(\ln\varrho_{\tau})^{\ell-n}\Lambda_{t_{\rm fin}}\left((\ln\varrho_{\rm in})^{n}\varrho_{\rm in}\right)\right]}&\nonumber \\
&&
\end{eqnarray}
where $\varrho_{\rm in} \equiv \sum_{k}p(a_{k}^{\rm in})\Pi_{k}^{\rm in}$ and $\varrho_{\tau} \equiv \sum_{m}p(a_{m}^{\rm fin})\Pi_{m}^{\rm fin}$ are, respectively, the ensemble average of the quantum system after the $1$st and $2$nd measurement of the TPM scheme. The validity of \eqref{moments} can be easily shown starting from the definition of the stochastic variable $\Delta\sigma$. Indeed, one gets
\begin{widetext}
\begin{align}
  \langle\Delta\sigma^{\ell}\rangle &=\sum_{m,k} p(a_{k}^{\rm in},a_{m}^{\rm fin})\left[\Delta\sigma(a^{\textrm{fin}}_{m},a^{\textrm{in}}_{k})\right]^{\ell}=
  \sum_{m,k} {\rm Tr}\left[\Pi^{\rm fin}_{m}\Lambda_{t_{\rm fin}}(\Pi^{\rm in}_{k})\right]p(a^{\rm in}_{k}) \left[\ln p(a^{\textrm{in}}_{k}) - \ln p(a^{\textrm{fin}}_{m})\right]^{\ell}= \nonumber \\
  &= \sum_{n=0}^{\ell}(-1)^{\ell-n}\binom{\ell}{n} {\rm Tr}\left[\sum_m \Pi^{\rm fin}_{m}\left(\ln p(a^{\textrm{fin}}_{m})\right)^{\ell-n}\Lambda_{t_{\rm fin}}\left(\sum_k \Pi^{\rm in}_{k}p(a^{\rm in}_{k})(\ln p(a^{\textrm{in}}_{k}))^{n}\right)\right] = \nonumber \\
  &= \sum_{n=0}^{\ell}(-1)^{\ell-n}\binom{\ell}{n}{\rm Tr}\left[(\ln\varrho_{\tau})^{\ell-n}\Lambda_{t_{\rm fin}}\left((\ln\varrho_{\rm in})^{n}\varrho_{\rm in}\right)\right].
\end{align}
\end{widetext}


\end{document}